\RequirePackage{fix-cm}
\documentclass[twocolumn]{svjour3}          % twocolumn
\smartqed  % flush right qed marks, e.g. at end of proof
\usepackage{graphicx}
\usepackage[export]{adjustbox}
\begin{document}

\title{Broadband Photocurrent Spectroscopy and Temperature Dependence of Band-gap of Few-Layer Indium Selenide}%\thanks{Grants or other notes about the article that should go on the front page should be placed here. General acknowledgments should be placed at the end of the article.}
%\subtitle{Do you have a subtitle? If so, write it here}


\author{Prasanna D. Patil   \and
        Milinda Wasala      \and
        Sujoy Ghosh         \and    
        Sidong Lei          \and
        Saikat Talapatra
}

\institute{P. D. Patil \at
              Department of Physics, Southern Illinois University, Carbondale, IL 62901, USA.\\
              \email{prasanna@siu.edu}
           \and
           M. Wasala \at
              Department of Physics, Southern Illinois University, Carbondale, IL 62901, USA.\\
              \emph{Present address: Science Department, Great Basin College, Elko, NV 89801, USA.}
           \and
           S. Ghosh \at
              Department of Physics, Southern Illinois University, Carbondale, IL 62901, USA.
           \and
           S. Lei \at
              Department of Physics and Astronomy, Georgia State University, Atlanta, GA 30303, USA.
           \and
           S. Talapatra \at
              Department of Physics, Southern Illinois University, Carbondale, IL 62901, USA.\\
              \email{saikat@siu.edu}           %  \\
}

\date{Received: date / Accepted: date}

\maketitle

\begin{abstract}
Understanding broadband photoconductive behaviour in two dimensional layered materials are important in order to utilize them for a variety of opto-electronic applications. Here we present our results of photocurrent spectroscopy measurements performed on few layer Indium Selenide (InSe) flakes. Temperature (T) dependent (40 K $<$ T $<$ 300 K) photocurrent spectroscopy was performed in order to estimate the band-gap energies $E_g(T)$ of InSe at various temperatures. Our measurements indicate that room temperature $E_g$ value for InSe flake was $\sim$ 1.254 eV, which increased to a value of $\sim$ 1.275 eV at low temperatures. The estimation of Debye temperatures by analysing the observed experimental variation of $E_g$ as a function of T using several theoretical models is presented and discussed.

\end{abstract}

\section{Introduction}
\label{intro}
Since the discovery of Graphene \cite{Graphene-2004}, it has been predicted that there could be numerous layered materials that can be isolated to a single or few layer form. These include transition metal dichalcogenides (TMDCs), metal oxides and single-element materials such as Silicene and Phosphorene \cite{NatureNews2D}. Isolating thin layers of these materials from their bulk counterpart imparts them  with exotic properties that could potentially lead to several applications \cite{wasala2017recent,Ghosh_2017_2D-Mat,Patil-2019-ACSNano-MIT,Patil-Electronics-2019,Ghosh_2018-Nanotech}. Specifically, it is predicted that several thin two dimensional (2D) layered semiconductors could possibly lead to multi functional opto-electronic applications due to their exotic photo electronic properties \cite{PRADHAN-BookChapter,C5CS00106D,LowDimPhotogating}. Initial studies, which favoured such predictions,  heavily investigated Molybdenum (Mo) and Tungsten (W) based binary chalcogenides \cite{2DTungstenChalcogenides,2D-TMDCsReview}, since, it was found that 2D MoS$_2$ can exhibit a photoresponsivity as high as  880 AW$^{-1}$ \cite{Lopez-Sanchez2013} and multilayer WS$_2$ can act as good photosensors \cite{WS2Nestor}.

\begin{figure*}
  \includegraphics[width=1\textwidth,center]{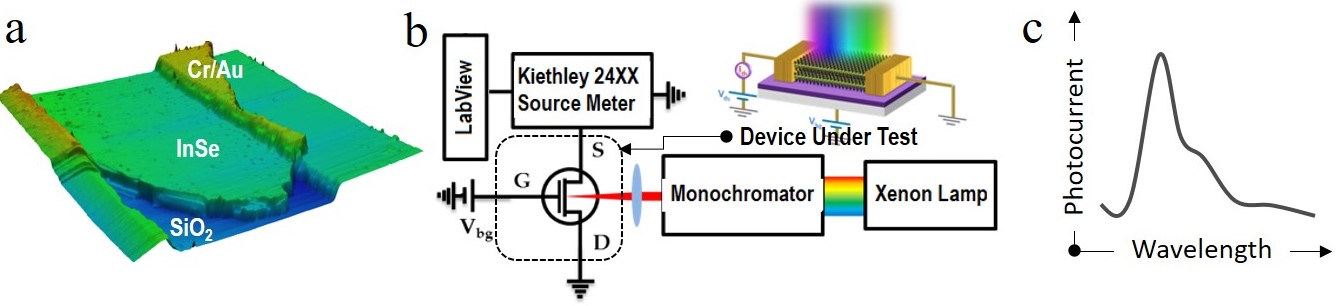}
\caption{(a) Atomic force microscopy (AFM) imags of device measured. (b) Schematic of experiential set-up that was used for measuring photocurrent spectroscopy (c) Typical photocurrent spectroscopy data (photocurrent as a function of wavelength).}
\label{Schematic}
\end{figure*}

Among many of the 2D materials that were initially investigated, several of them, which includes MoS$_2$, WS$_2$ etc., are direct bandgap materials in single layer form. The possibility of a single atomic layer to be viable for key optoelctronic processes that demands enhanced optical absorption \cite{Photo-FETs}, therefore, seems unlikely. As such, 2D layered materials which possess direct bandgap in its few layered form, could perhaps provide ample optical absorption, leading to a variety of suitable application, such as high sensitivity photo detectors, photo switch, active materials for solar cells etc \cite{Photo-FETs}. Several group III-VI layered compounds belong to this category in which direct band-gap persists even if the material is few layers thick \cite{wasala2017recent}. These class of materials which includes InSe, show promising electronic and optoelectronic properties \cite{wasala2017recent,C5CS00106D}.These Se based system due to its high photoresponsivity and wide spectral response can be used for photovoltaics and photodetector applications. InSe was reported to have a small band gap of 1.3 eV having a broadband spectral response \cite{InSeTamalampudi,Mudd-AdvMat-2013}. Investigators have also shown broad spectral response as well as high performance of flexible photodetectors using few layes of InSe \cite{InSeTamalampudi,SidongACSNanoInSe}. These initial investigations seems very promising and indicates the prospect of InSe to become one of the choice materials for wide variety of opto-electronics applications. Thus, understanding the spectral response of these materials are important both from fundamental and technical point of view. Here we present the photocurrent spectral response of few-layered InSe flake over a wide range of temperature (40 K $<$ T $<$ 300 K). Room temperature band-gap value (E$_g$) for InSe flake was found to $\sim$ 1.254 eV. Values of E$_g$ showed a slow increase with decrease in temperature (up to $\sim$ 120 K). For T below 120 K, the dependence of E$_g$ with T was found to be weak. Analysis of the variation of E$_g$ as a function T was performed using several theoretical models such as Bose-Einstein function \cite{Pejova2010Temp}, Double Bose-Einstein function \cite{PhysRevB.86.195208} and Manoogian-Leclerc equation \cite{Manoogian-Leclerc_1979} in order to understand the effects of electron - phonon interaction as well as lattice dilation. From these fittings, we anticipate that the Debye temperature ($\Theta_D$) of InSe flakes $\sim$ 260 K.

%\paragraph{Paragraph headings} Use paragraph headings as needed.
\section{Materials and methods}
\label{Materials}
\subsection{Indium Selenide crystal synthesis and exfoliation}

Few layer InSe devices were fabricated from thin flakes obtained through mechanical exfoliation of bulk crystal grown using thermal treatment of a nonstoichiometric mix of Indium ($>$ 99.99\%, Alfa Aesar Co.) and Selenium ($>$ 99.99\%, Sigma-Aldrich Co.) with a molar ratio of 52 : 48 in a sealed quartz tube under millitorr vacuum \cite{SidongACSNanoInSe}. At the beginning, the system was heated to 685$^0$C. This temperature was maintained for few hours to ensure a complete reaction between In and Se. Thereafter, the temperature was raised to 700$^0$C and was maintained as this temperature of another 3 hours. The system was then cooled down to 500$^0$C at 10$^0$C per hour rate. Once the system was cooled down to 500$^0$C, it was allowed to cool down naturally to the room temperature to acquire high quality InSe crystals. 

\subsection{Device fabrication}
The detail device fabrication method and electrical transport characterization are presented in one of our previous publications \cite{WasalaInSe_2020,wasala2021-OxfMat}. In short the devices were fabricated on Silicon/Silicon dioxide (SiO$_2$) wafers with a (SiO$_2$) thickness of 1000 nm. A typical device fabrication routine involved evaporating metal contacts on top of a suitable flake residing on the Si/SiO$_2$ wafer through shadow masking. We have used either nickel or gold transmission electron microscopy grids for the masking process. After securing TEM grid carefully on top of the flake, system was mounted inside the thermal evaporator chamber. The chamber was then pumped down to  10$^{-6}$ torr and was held at this pressure overnight. A thin layer of Chromium (Cr) and gold (Au) electrodes were deposited through the mask by evaporating them using a tungsten (W) boat. Typically, about 40 nm of Cr layers and about 160 nm thick Au layer was deposited. After metal deposition, the system was cooled down to room temperature before the devices could be taken out from the metal deposition chamber. The height profile of the device was measured using the contact mode Atomic Force Microscopy (AFM).  Height of the device discussed below measured as $\sim$ 34 nm, which is corresponds to 40 InSe layers. Also, active device area was calculated as $2.1 \times 10^{-9} m^{2}$.  The metal contacted flakes were then placed on a ceramic chip holder and gold connecting wires were bonded to the metal contacts using a wire bonder. The chip was then mounted on a closed cycle helium cryostat for with an optical window for desired opto-electronic measurements. 

\begin{figure*}
  \includegraphics[width=0.75\textwidth,center]{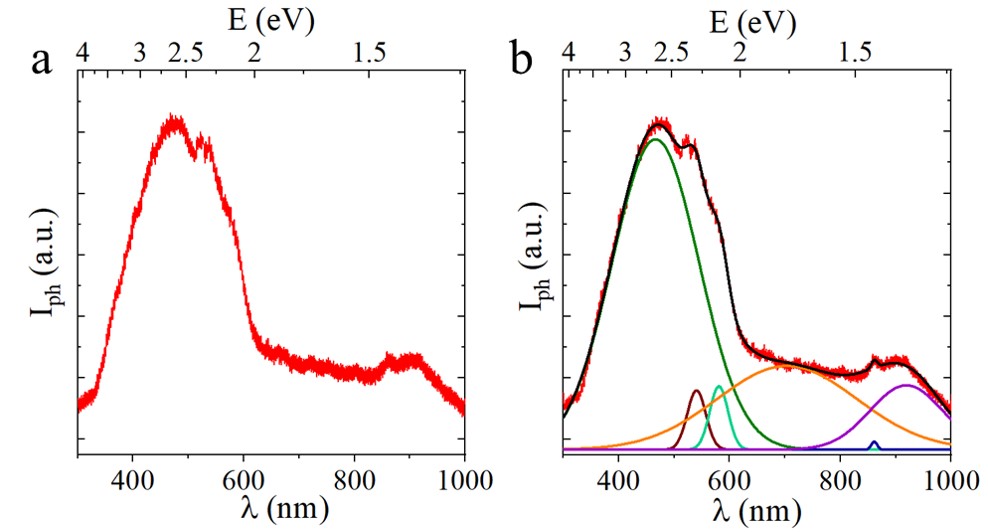}
\caption{(a) Room temperature photocurrent spectra of InSe device. (b) Fitted spectra with multiple Gaussian peaks}
\label{Spectra}
\end{figure*}

\subsection{Photocurrent spectroscopy measurements}
Broadband Photo-electronic conduction (PC) characterization was performed using an in house built photocurrent spectroscopy measurement set-up. Xenon lamp was used as a light source to generate broad spectrum of wavelengths (300 nm $\leq \lambda \leq $ 1000 \ nm) and was passed through a Monochromator. These monochromatic light sources were guided on to the device using an optical wave-guide and and focused on to the device using a convex lens with a focal length of 15 cm. Kiethley 2400 series source meters controlled with in-house developed LabVIEW module was utilized for measuring the photo response. Schematic diagram of the experimental setup and typical data obtained is shown in Fig. \ref{Schematic}.  

\section{Results and discussions}
\label{Result}
Photocurrent spectroscopy, in the past, have been used widely in order to estimate the broadband optical absorption behavior of a variety of semiconductor material \cite{TMDc1982-Kam,SciRep2D-Klots2014} including,  InSe\cite{Mudd-AdvMat-2013,SidongACSNanoInSe}, MoS$_2$\cite{CommPhys-Vaquero2020}, MoSe$_2$\cite{2DMater_MoSe2-2017} etc. The generation and/or variation of current as a function of wavelength of light radiation impinging on the material is the core information that can be obtained using this technique. The schematics of our set up and the data obtained using this setup is shown in Fig. 1 and Fig. 2 respectively. The data shown in Fig. 2a, was obtained from a 34 nm thick InSe flake (Fig 1a). From fig. 2a, it is clear that the InSe samples show a broadband photoconductivity, with prominent photoconductivity peaks at several wavelengths. In order to gain insight into the various peaks present in the data, we have have utilized peak fitting with appropriate deconvolution. We found that the peaks from spectra, (Fig. \ref{Spectra}a) can be deconvoluted using Eq. \ref{Gaussian}.

\begin{equation}
\label{Gaussian}
I_{ph}(\lambda)=\frac{A}{\sigma \sqrt{2\pi}} e^{-\frac{1}{2}(\frac{\lambda-\lambda_0}{\sigma})^2}
\end{equation}
where A is constant, $\lambda_0$ is peak position and $\sigma$ is standard deviation. Values of $\lambda_0$ and $\sigma$ for all five peaks from Fig. \ref{Spectra}b are listed in Table \ref{PeakGaussian}.

\begin{figure*}
  \includegraphics[width=0.66\textwidth,center]{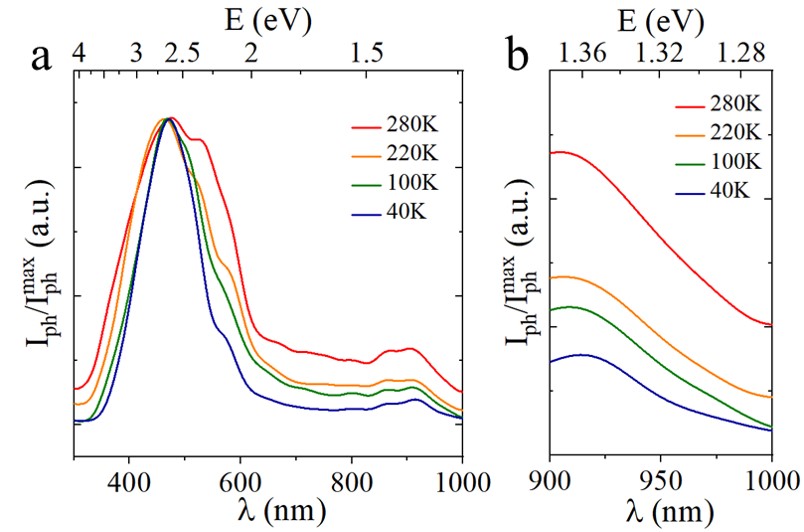}
\caption{(a) Temperature dependent photocurrent spectra of InSe device. (b) spectra obtained at longer wavelength showing band edge used for calculating band-gap in Fig. \ref{BandGap}}
\label{SpectraTemp}
\end{figure*}

The general behavior of the photoconductivity as a function of wavelength is presented in figure 2. From figure fig. 2a, it can be seen that the photocurrent slowly increases for decreasing wavelengths close to the bandgap (or E $>$ 1.26 eV), and shows a broad peak around 2.5eV. Similar photocurrent peaks are reported for other photoconductive materials in the past \cite{JAP-1977-HgI2,JAP-Bube-1987-HgI2,Burshtein-JAP-1986-HgI2,HgI2-BookChapter} and can be explained as follows. Initially, the photocurrent increases slowly and then starts to increase sharply from $\sim$ 2 eV till $\sim$ 2.66 eV. This could be generally attributed to increased photon absorption, which leads to an increase in the generated carriers. As the wavelength decreases further the carriers generated increases (at higher energies). The carriers generated at higher energies (lower wavelengths) tends to recombine with surface states, and therefore do not contribute to the photocurrent and hence at lower wavelengths photocurrent decreases sharply (as seen in fig. 2a for E $>$ 2.5 eV). 

In one of the previous investigations \cite{SidongACSNanoInSe} related to photocurrent spectroscopy measurement of similar InSe samples (measured within the wavelength of interval 400 nm $<$ $\lambda$ $<$ 800 nm) it was noted that, in case of InSe, electronic transition from P$_{x,y}$-like orbital to the bottom of conduction band will occur between 2.25 ev (550 nm) and 3.1 eV (400 nm) and electronic transition from P$_z$-like orbital to the bottom of conduction band will occur between 1.55 eV (800 nm) and 2.25 ev (550 nm) \cite{SidongACSNanoInSe}. Peaks occurring at similar energy values were observed for our sample as well, as seen from the deconvoluted peaks. We see these similar peaks as indicated in Table 1. For $\lambda$ between 800 nm - 1000 nm, a broad shoulder around 900nm ($\sim$1.4 eV;  deconvoluted to two peaks) as seen in figure 2b. Generally, the higher wavelength peaks observed near the band edge are attributed to deep level impurities \cite{InSe-Film-1987}, we believe such impurities in our samples are responsible for the peaks observed in the higher wavelength region.

\begin{table}
\centering
\caption{Parameter for peak fitting for photocurrent spectra. Here, $\lambda_0$ is peak position and $\sigma$ is standard deviation of Gauss distribution (Eq. \ref{Gaussian})}
\label{PeakGaussian}
\begin{tabular}{llll}
\hline\noalign{\smallskip}
Peak \# & $\lambda_0$ (nm) & $\sigma$ & E ($\lambda_0$) (eV) \\
\noalign{\smallskip}\hline\noalign{\smallskip}
Peak 1 & 466.56 & 78.09  & 2.658\\
Peak 2 & 540.40 & 16.77 & 2.296\\
Peak 3 & 580.79 & 16.55 & 2.134\\
Peak 4 & 703.68 & 125.89 & 1.772\\
Peak 5 & 861.10 & 5.26 & 1.439\\
Peak 6 & 918.94 & 65.53 & 1.349\\
\noalign{\smallskip}\hline
\end{tabular}
\end{table}

\subsection{Band gap determination from photocurrent spectroscopy}

In order to estimate the band gap from photocurrent spectra, we have utilized methods similar to extracting band gap from UV-Vis absorption spectroscopy. Since, the photocurrent generated in a semiconductor is proportional to the number of absorbed photons and the absorption coefficient ($\alpha$) of a direct bandgap semiconductor is proportional to the square root of the difference of the photon energy $(E_{ph})$ and the bandgap energy, $(E_g)$, or in other words $\alpha$ $\sim$ $(E_{ph} - E_g)^{0.5}$. Therefore, it can be assumed that $(I_{ph}) \sim (E_{ph} - E_g)^{0.5}$ as well and one can generate a plot similar to Tauc plot \cite{Tauc-1966} for extracting the band gap values. In the past several investigators have utilized the aforementioned technique and used photocurrent spectroscopy for extraction and estimation of band gap \cite{SidongACSNanoInSe,2DMater_MoSe2-2017,SciRep2D-Klots2014,TMDc1982-Kam}. For this purpose we have plotted the quantity $[I_{ph} \times E]^{2}$ as a function of $E$ as shown in fig 4. For estimation of the band gap from this plot was performed by extrapolating the straight line portion of the data to x-axis intercept. In Fig 4. we have presented such extrapolations for various temperatures. The room temperature band gap value for the measured sample was found to be $\sim$ 1.254eV. This value is similar to the values obtained for multi layered InSe flakes from past investigations \cite{InSeTamalampudi,Mudd-AdvMat-2013,InSe-Bandgap-Physica-2014,Temp-Bandgap-InSe-2020}. 

\begin{figure*}
  \includegraphics[width=0.85\textwidth,center]{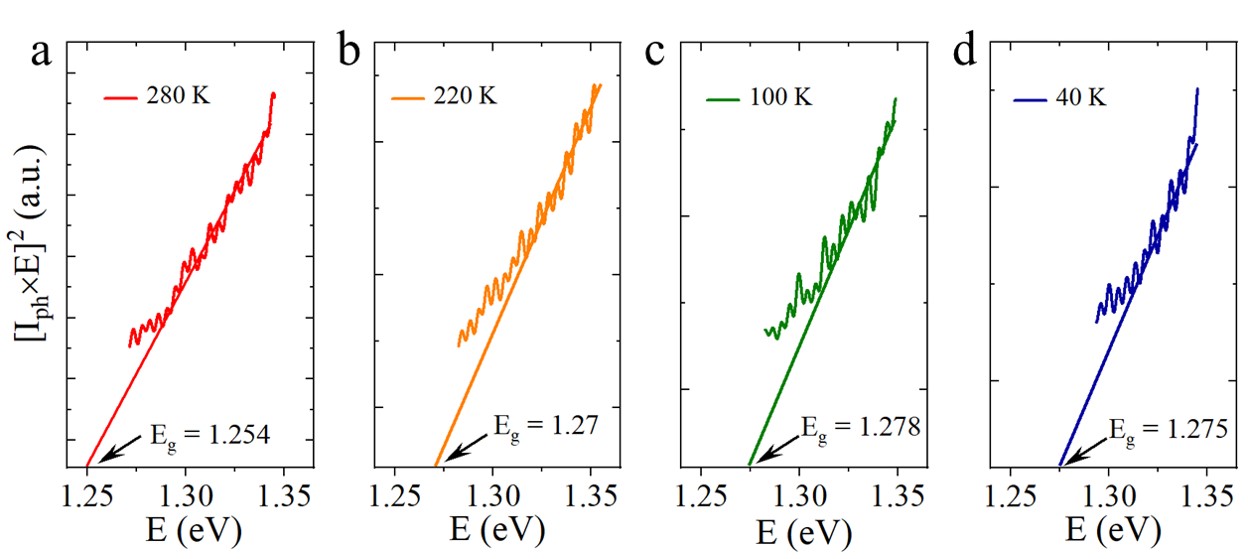}
\caption{Band gap extraction using extrapolation ([I$_{ph} \times$ E]$^2$  \textit{vs.} E) at (a) 280 K, (b) 220 K, (c) 100 K and (d) 40 K}
\label{BandGap}
\end{figure*}

\subsection{Temperature dependence of band-gap}

\begin{figure*}
  \includegraphics[width=1\textwidth,center]{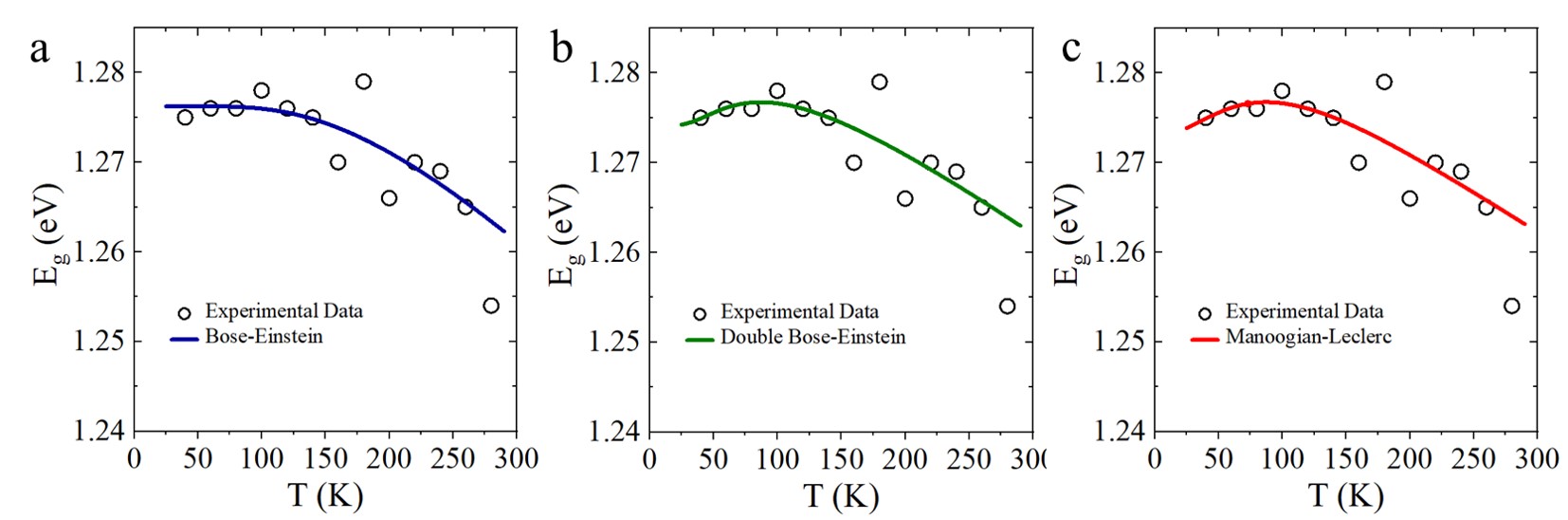}
\caption{Temperature dependence of the band-gap energy of the InSe, fitted to (a) Bose-Einstein function (blue), (b) Double Bose-Einstein function (green) and (c) Manoogian-Leclerc equation (red).}
\label{BandGapTemp}
\end{figure*}

Temperature (T) dependence of $E_g$ derived from photocurrent spectroscopy of InSe (Fig. \ref{BandGap}) is presented in Fig. \ref{BandGapTemp}. The information pertaining to the variation of $E_g$ in semiconductors as a function of temperature is of significance since, such information can be analyzed in order to have fundamental insights about core materials property. Generally, the variation of $E_g$ as a function of T is caused due to the combination of electron-phonon interaction and lattice dilation. In the past, several semi-empirical equations \cite{VARSHNI_1967,Manoogian-Leclerc_1979,Pejova2010Temp,PhysRevB.86.195208} are developed in order fit the experimental data obtained. One of the most popular approach is to fit the experimental data with Varshani's equation \cite{VARSHNI_1967}, however, it has been argued that in some cases where lattice dilation effects could be significant, a good fit to the experimental data using Varshni equation was not achievable \cite{CanJPhys_1984}. Several other equations for example; Bose-Einstein (BE) function \cite{Pejova2010Temp}, Double Bose-Einstein (DBE) function \cite{PhysRevB.86.195208} and Manoogian - Leclerc (ML) \cite{Manoogian-Leclerc_1979} equation are also used widely to fit the variation of $E_g$ as a function of T. Below we briefly describe each of these equations and discuss in detail the results obtained by fitting them to our data.

Bose-Einstein function \cite{Pejova2010Temp} takes into consideration the interaction between electron and phonon and can be represented as shown in Eq. (\ref{BoseEist}).
\begin{equation}
\label{BoseEist}
E_g(T)=E_g(0)-\frac{2a_B}{exp\big(\frac{\Theta_E}{T}\big)-1}
\end{equation}
In this equation, $E_g(0)$ is a band gap at absolute zero (0 K), $a_B$ is a measure of strength of the electron-phonon interaction coupling within the crystal and $\Theta_E$ is the Einstein characteristic temperature. $\Theta_E$ is defined as the average temperature of the phonons which are interacting with the electrons. It is deduced that Debye phonon spectrum with the Debye temperature $\Theta_D$ is equivalent to an Einstein oscillator with a temperature $\Theta_E$, with $\Theta_D = 4/3 \times \Theta_E$ \cite{Pejova2010Temp}. From fitting of Eq. (\ref{BoseEist}) in Fig. \ref{BandGapTemp}, a value of $\Theta_D$ is extracted to be 782.2 K. This value of $\Theta_D$ is significantly higher than theoretical value of $\Theta_D$ = 190 K \cite{InSeDebyeTempBook} as well as experimentally determined maximum possible $\Theta_D$ of 275 $\pm 15$ K for layered InSe \cite{InSeDebye-Tyurin2007}. One reason for such discrepancy could arise from the fact that Bose-Einstein function uses only one Einstein oscillator. In presence of one Einstein oscillator, band-gap energy will either monotonically increase or saturate at constant value as temperature T $\to$ 0 \cite{PhysRevB.86.195208}. For our experimental data, we found a small decrease in ban-gap energy at low temperatures (T $\le$ 100 K). This behaviour could be elucidate by considering a contribution from low energy (secondary) phonon with opposite weight, leading to decrease in band-gap energy for T $\to$ 0 \cite{PhysRevB.86.195208}. Such contributions can be incorporated by using Double Bose-Einstein function of the form shown in Eq. \ref{DoubleBoseEist} was used for fitting.
\begin{equation}
\label{DoubleBoseEist}
E_g(T)=E_g(0)-\frac{2a_{B1}}{exp\big(\frac{\Theta_{E1}}{T}\big)-1}+\frac{2a_{B2}}{exp\big(\frac{\Theta_{E2}}{T}\big)-1}
\end{equation}
where $E_g(0)$, $a_{B1}$, $a_{B2}$, $\Theta_{E1}$ and $\Theta_{E1}$ have same meaning as previously described. The second Einstein oscillator (third term in Eq. \ref{DoubleBoseEist} carries a opposite weight to that of the first oscillator. From fitting of Eq. (\ref{DoubleBoseEist}) in Fig. \ref{BandGapTemp}, a value of $\Theta_D$ is extracted to be 261.3 K, which is much closer to experimentally determined maximum possible Debye temperature of 275 $\pm 15$ K for layered InSe \cite{InSeDebye-Tyurin2007}. The frequencies of the two Bose-Einstein oscillators correspond to optic and acoustic phonons \cite{PhysRevB.86.195208}.

Past investigations have also shown that consideration of lattice dilation effects are crucial in order to explain the temperature variation of band gaps in a variety of semiconductors \cite{Manoogian-Leclerc_1979}. In order to determine if there is any lattice dilation effect along with electron-phonon interactions that might influence the variation of $E_{g}$ with T in our sample, we also considered fitting our data to Manoogian-Leclerc equation, which incorporates the lattice dilation effect term. Eq. \ref{MangLec} shows the form of Manoogian-Leclerc equation.
\begin{equation}
\label{MangLec}
E_g(T)=E_g(0)-UT^s-V\Theta_E \bigg[ coth\bigg(\frac{\Theta_{E}}{2T}\bigg)-1\bigg]
\end{equation}
In Eq. \ref{MangLec}, $E_g(0)$ is a band gap at absolute zero (0 K) and U, V, and s are temperature-independent constants. Second and third terms on right hand side of equation correspond to lattice dilation term and electron-phonon interactions respectively. From fitting of Eq. (\ref{DoubleBoseEist}) in Fig. \ref{BandGapTemp}c, a value of $\Theta_D$ is extracted to be 403.5 K. This value of $\Theta_D$ is substantially higher than experimental values of $\Theta_D$ previously reported for layered InSe \cite{InSeDebye-Tyurin2007}.  Observing the $\Theta_D$ values obtained specifically from all the fits and comparing it to typical values of $\Theta_D$ reported for layered InSe \cite{InSeDebye-Tyurin2007}, we believe that Double Bose-Einstein function is the most closest model that can explain the bandgap shift as a function of temperature observed in our sample. Various parameters obtained from all the fits are summarized in Table \ref{ParameterTable}. From the information presented in Table \ref{ParameterTable}, we note that although the estimated values of $E_g(0)$ for all the models used to fit our experimental data yielded reasonably consistent value of $E_g(0)$ $\sim$ 1.27 eV, this was not the case for values obtained for $\Theta_D$. One reason for this could be that $\Theta_D$ for a material is often dependent on multiple factors involving the thermodynamic and mechanical properties of material \cite{UllrichAIP2017} and hence it becomes extremely difficult to have an accurate and consistent estimation of this value.   

\begin{table}
\centering
\caption{Values of parameters obtained by fitting of the experimental band-gap energy data. Here $E_g(0)$ is band-gap at 0 K, $\Theta$ is Einstein Temperature and  $\Theta_D$ is Debye temperature. In case of double Bose-Einstein function, 1 and 2 correspond to Einstein oscillators associated with electron interacting with primary and secondary phonons respectively.}
\label{ParameterTable}% Give a unique label For LaTeX tables use
\begin{tabular}{llll}
\hline\noalign{\smallskip}
fitting & $E_g(0)$ & $\Theta$ & $\Theta_D$ \\
        & (eV)     &    (K)     &    (K) \\
\noalign{\smallskip}\hline\noalign{\smallskip}
Bose-Einstein & 1.276 & 588.1 & 782.2\\
Double Bose-Einstein & 1.274 & $^1$196.5 & $^1$261.3\\
                     &       & $^2$180.4 & $^2$239.9\\
Manoogian-Leclerc & 1.272 & 303.4 & 403.5 \\
\noalign{\smallskip}\hline
\end{tabular}
\end{table}

\section{Conclusion}

In summary, broadband photoconductive behavior was observed in 34 nm thick flakes of layered InSe. We have tried to investigate the variation of band gap as a function of temperature by performing photocurrent spectroscopy measurements. The data obtained was used to estimate Debye temperature ($\Theta_D$) values of InSe using established theoretical models. The $\Theta_D$ of $\sim$ 260 K obtained using Double Bose-Einstein equation is similar to the values obtained for this quantity for layered InSe. This strongly suggests that Double Bose-Einstein equation is perhaps the best model for explaining the variation of band gap of thin InSe flakes. This would mean (a) that band-gap shifts observed in few layer InSe samples presented in this study depends on electron-phonon interactions, which includes low energy secondary phonons and (b) lattice dilation effects are perhaps negligible in our samples. Most importantly, the variation of the band gap of InSe with temperature presented here is extremely valuable for strengthening our fundamental understanding needed for developing a variety of technological applications using optically active 2D materials.

\begin{acknowledgements}
This work was supported by the U.S. Army Research Office MURI grant \#W911NF-11-1-0362. S.T. and P.D.P. acknowledges the support from Indo-U.S. Virtual Networked Joint Center Project on “Light Induced Energy Technologies: Utilizing Promising 2D Nanomaterials (LITE UP 2D)” through the grant number IUSSTF/JC-071/2017. M.W. and P.D.P acknowledges the College of Science Dissertation Research Award and Graduate School Doctoral Fellowship respectively, awarded at Southern Illinois University Carbondale (SIUC).
\end{acknowledgements}

\section*{Conflict of interest}
The authors declare that they have no conflict of interest.


%\begin{thebibliography}{}
% and use \bibitem to create references. Consult the Instructions for authors for reference list style.
% etc
%\end{thebibliography}

\bibliographystyle{unsrt}
\bibliography{Ref.bib}

\begin{thebibliography}{10}

\bibitem{Graphene-2004}
K.~S. Novoselov, A.~K. Geim, S.~V. Morozov, D.~Jiang, Y.~Zhang, S.~V. Dubonos,
  I.~V. Grigorieva, and A.~A. Firsov.
\newblock Electric field effect in atomically thin carbon films.
\newblock {\em Science}, 306(5696):666--669, 2004.

\bibitem{NatureNews2D}
Elizabeth Gibney.
\newblock The super materials that could trump graphene.
\newblock {\em Nature}, 522:274–276, Jun 2015.

\bibitem{wasala2017recent}
Milinda Wasala, Hansika~I Sirikumara, Yub~Raj Sapkota, Stephen Hofer, Dipanjan
  Mazumdar, Thushari Jayasekera, and Saikat Talapatra.
\newblock Recent advances in investigations of the electronic and
  optoelectronic properties of group iii, iv, and v selenide based binary
  layered compounds.
\newblock {\em Journal of Materials Chemistry C}, 5(43):11214--11225, 2017.

\bibitem{Ghosh_2017_2D-Mat}
Sujoy Ghosh, Prasanna~D Patil, Milinda Wasala, Sidong Lei, Andrew Nolander,
  Pooplasingam Sivakumar, Robert Vajtai, Pulickel Ajayan, and Saikat Talapatra.
\newblock Fast photoresponse and high detectivity in copper indium selenide
  (cuin7se11) phototransistors.
\newblock {\em 2D Materials}, 5(1):015001, oct 2017.

\bibitem{Patil-2019-ACSNano-MIT}
Prasanna~D. Patil, Sujoy Ghosh, Milinda Wasala, Sidong Lei, Robert Vajtai,
  Pulickel~M. Ajayan, Arindam Ghosh, and Saikat Talapatra.
\newblock Gate-induced metal–insulator transition in 2d van der waals layers
  of copper indium selenide based field-effect transistors.
\newblock {\em ACS Nano}, 13(11):13413--13420, 2019.

\bibitem{Patil-Electronics-2019}
Prasanna~D. Patil, Sujoy Ghosh, Milinda Wasala, Sidong Lei, Robert Vajtai,
  Pulickel~M. Ajayan, and Saikat Talapatra.
\newblock Electric double layer field-effect transistors using two-dimensional
  (2d) layers of copper indium selenide (cuin7se11).
\newblock {\em Electronics}, 8(6):645, 2019.

\bibitem{Ghosh_2018-Nanotech}
Sujoy Ghosh, Milinda Wasala, Nihar~R Pradhan, Daniel Rhodes, Prasanna~D Patil,
  Michael Fralaide, Yan Xin, Stephen~A McGill, Luis Balicas, and Saikat
  Talapatra.
\newblock Low temperature photoconductivity of few layer p-type tungsten
  diselenide ({WSe}2) field-effect transistors ({FETs}).
\newblock {\em Nanotechnology}, 29(48):484002, oct 2018.

\bibitem{PRADHAN-BookChapter}
Nihar~R. Pradhan, Rukshan Thantirige, Prasanna~D. Patil, Stephen~A. McGill, and
  Saikat Talapatra.
\newblock 6 - electronic and optoelectronic properties of the heterostructure
  devices composed of two-dimensional layered materials.
\newblock In Satyabrata Jit and Santanu Das, editors, {\em 2D Nanoscale
  Heterostructured Materials}, Micro and Nano Technologies, pages 151--193.
  Elsevier, 2020.

\bibitem{C5CS00106D}
Michele Buscema, Joshua~O. Island, Dirk~J. Groenendijk, Sofya~I. Blanter,
  Gary~A. Steele, Herre S.~J. van~der Zant, and Andres Castellanos-Gomez.
\newblock Photocurrent generation with two-dimensional van der waals
  semiconductors.
\newblock {\em Chem. Soc. Rev.}, 44:3691--3718, 2015.

\bibitem{LowDimPhotogating}
Hehai Fang and Weida Hu.
\newblock Photogating in low dimensional photodetectors.
\newblock {\em Advanced Science}, 4(12):1700323, 2017.

\bibitem{2DTungstenChalcogenides}
Melinda Mohl, Anne-Riikka Rautio, Georgies~Alene Asres, Milinda Wasala,
  Prasanna~Dnyaneshwar Patil, Saikat Talapatra, and Krisztian Kordas.
\newblock 2d tungsten chalcogenides: Synthesis, properties and applications.
\newblock {\em Advanced Materials Interfaces}, 7(13):2000002, 2020.

\bibitem{2D-TMDCsReview}
Song-Lin Li, Kazuhito Tsukagoshi, Emanuele Orgiu, and Paolo Samorì.
\newblock Charge transport and mobility engineering in two-dimensional
  transition metal chalcogenide semiconductors.
\newblock {\em Chem. Soc. Rev.}, 45:118--151, 2016.

\bibitem{Lopez-Sanchez2013}
Oriol Lopez-Sanchez, Dominik Lembke, Metin Kayci, Aleksandra Radenovic, and
  Andras Kis.
\newblock Ultrasensitive photodetectors based on monolayer mos2.
\newblock {\em Nature Nanotechnology}, 8(7):497--501, Jul 2013.

\bibitem{WS2Nestor}
Néstor Perea-López, Ana~Laura Elías, Ayse Berkdemir, Andres Castro-Beltran,
  Humberto~R. Gutiérrez, Simin Feng, Ruitao Lv, Takuya Hayashi, Florentino
  López-Urías, Sujoy Ghosh, Baleeswaraiah Muchharla, Saikat Talapatra,
  Humberto Terrones, and Mauricio Terrones.
\newblock Photosensor device based on few-layered ws2 films.
\newblock {\em Advanced Functional Materials}, 23(44):5511--5517, 2013.

\bibitem{Photo-FETs}
Dominik Kufer and Gerasimos Konstantatos.
\newblock Photo-fets: Phototransistors enabled by 2d and 0d nanomaterials.
\newblock {\em ACS Photonics}, 3(12):2197--2210, 2016.

\bibitem{InSeTamalampudi}
Srinivasa~Reddy Tamalampudi, Yi-Ying Lu, Rajesh~Kumar U., Raman Sankar, Chun-Da
  Liao, Karukanara~Moorthy B., Che-Hsuan Cheng, Fang~Cheng Chou, and Yit-Tsong
  Chen.
\newblock High performance and bendable few-layered inse photodetectors with
  broad spectral response.
\newblock {\em Nano Letters}, 14(5):2800--2806, 2014.
\newblock PMID: 24742243.

\bibitem{Mudd-AdvMat-2013}
Garry~W. Mudd, Simon~A. Svatek, Tianhang Ren, Amalia Patanè, Oleg Makarovsky,
  Laurence Eaves, Peter~H. Beton, Zakhar~D. Kovalyuk, George~V. Lashkarev,
  Zakhar~R. Kudrynskyi, and Alexandr~I. Dmitriev.
\newblock Tuning the bandgap of exfoliated inse nanosheets by quantum
  confinement.
\newblock {\em Advanced Materials}, 25(40):5714--5718, 2013.

\bibitem{SidongACSNanoInSe}
Sidong Lei, Liehui Ge, Sina Najmaei, Antony George, Rajesh Kappera, Jun Lou,
  Manish Chhowalla, Hisato Yamaguchi, Gautam Gupta, Robert Vajtai, Aditya~D.
  Mohite, and Pulickel~M. Ajayan.
\newblock Evolution of the electronic band structure and efficient
  photo-detection in atomic layers of inse.
\newblock {\em ACS Nano}, 8(2):1263--1272, 2014.
\newblock PMID: 24392873.

\bibitem{Pejova2010Temp}
Biljana Pejova, Bahattin Abay, and Irina Bineva.
\newblock Temperature dependence of the band-gap energy and sub-band-gap
  absorption tails in strongly quantized znse nanocrystals deposited as thin
  films.
\newblock {\em The Journal of Physical Chemistry C}, 114(36):15280--15291,
  2010.

\bibitem{PhysRevB.86.195208}
J.~Bhosale, A.~K. Ramdas, A.~Burger, A.~Mu\~noz, A.~H. Romero, M.~Cardona,
  R.~Lauck, and R.~K. Kremer.
\newblock Temperature dependence of band gaps in semiconductors:
  Electron-phonon interaction.
\newblock {\em Phys. Rev. B}, 86:195208, Nov 2012.

\bibitem{Manoogian-Leclerc_1979}
A.~Manoogian and A.~Leclerc.
\newblock Determination of the dilation and vibrational contributions to the
  energy band gaps in germanium and silicon.
\newblock {\em physica status solidi (b)}, 92(1):K23--K27, 1979.

\bibitem{WasalaInSe_2020}
Milinda Wasala, Prasanna~D Patil, Sujoy Ghosh, Rana Alkhaldi, Lincoln Weber,
  Sidong Lei, Robert Vajtai, Pulickel~M Ajayan, and Saikat Talapatra.
\newblock Influence of channel thickness on charge transport behavior of
  multi-layer indium selenide ({InSe}) field-effect transistors.
\newblock {\em 2D Materials}, 7(2):025030, feb 2020.

\bibitem{wasala2021-OxfMat}
Milinda Wasala, Prasanna~D. Patil, Sujoy Ghosh, Lincoln Weber, Sidong Lei, and
  Saikat Talapatra.
\newblock Role of layer thickness and field-effect mobility on
  photoresponsivity of indium selenide (inse) based phototransistors, 2021.

\bibitem{TMDc1982-Kam}
K.~K. Kam and B.~A. Parkinson.
\newblock Detailed photocurrent spectroscopy of the semiconducting group vib
  transition metal dichalcogenides.
\newblock {\em The Journal of Physical Chemistry}, 86(4):463--467, 1982.

\bibitem{SciRep2D-Klots2014}
A.~R. Klots, A.~K.~M. Newaz, Bin Wang, D.~Prasai, H.~Krzyzanowska, Junhao Lin,
  D.~Caudel, N.~J. Ghimire, J.~Yan, B.~L. Ivanov, K.~A. Velizhanin, A.~Burger,
  D.~G. Mandrus, N.~H. Tolk, S.~T. Pantelides, and K.~I. Bolotin.
\newblock Probing excitonic states in suspended two-dimensional semiconductors
  by photocurrent spectroscopy.
\newblock {\em Scientific Reports}, 4(1):6608, Oct 2014.

\bibitem{CommPhys-Vaquero2020}
Daniel Vaquero, Vito Cleric{\`o}, Juan Salvador-S{\'a}nchez, Adri{\'a}n
  Mart{\'i}n-Ramos, Elena D{\'i}az, Francisco Dom{\'i}nguez-Adame, Yahya~M.
  Meziani, Enrique Diez, and Jorge Quereda.
\newblock Excitons, trions and rydberg states in monolayer mos2 revealed by
  low-temperature photocurrent spectroscopy.
\newblock {\em Communications Physics}, 3(1):194, Oct 2020.

\bibitem{2DMater_MoSe2-2017}
Jorge Quereda, Talieh~S Ghiasi, Feitze~A van Zwol, Caspar~H van~der Wal, and
  Bart~J van Wees.
\newblock Observation of bright and dark exciton transitions in monolayer
  {MoSe}2 by photocurrent spectroscopy.
\newblock {\em 2D Materials}, 5(1):015004, oct 2017.

\bibitem{JAP-1977-HgI2}
F.~Adduci, A.~Cingolani, M.~Ferrara, M.~Lugarà, and A.~Minafra.
\newblock Photoelectromagnetic properties of hgi2.
\newblock {\em Journal of Applied Physics}, 48(1):342--345, 1977.

\bibitem{JAP-Bube-1987-HgI2}
Jonathan Bornstein and Richard~H. Bube.
\newblock Photoelectronic properties of hgi2.
\newblock {\em Journal of Applied Physics}, 61(7):2676--2678, 1987.

\bibitem{Burshtein-JAP-1986-HgI2}
Z.~Burshtein, Justin~K. Akujieze, and E.~Silberman.
\newblock Carrier surface generation and recombination effects in
  photoconduction of hgi2 single crystals.
\newblock {\em Journal of Applied Physics}, 60(9):3182--3187, 1986.

\bibitem{HgI2-BookChapter}
X.J. Bao, T.E. Schlesinger, and R.B. James.
\newblock Chapter 4 - electrical properties of mercuric iodide.
\newblock In T.E. Schlesinger and Ralph~B. James, editors, {\em Semiconductors
  and Semimetals}, volume~43 of {\em Semiconductors and Semimetals}, pages
  111--168. Elsevier, 1995.

\bibitem{InSe-Film-1987}
M.~Yudasaka, T.~Matsuoka, and K.~Nakanishi.
\newblock Indium selenide film formation by the double-source evaporation of
  indium and selenium.
\newblock {\em Thin Solid Films}, 146(1):65--73, 1987.

\bibitem{Tauc-1966}
J.~Tauc, R.~Grigorovici, and A.~Vancu.
\newblock Optical properties and electronic structure of amorphous germanium.
\newblock {\em physica status solidi (b)}, 15(2):627--637, 1966.

\bibitem{InSe-Bandgap-Physica-2014}
Bekir Gürbulak, Mehmet Şata, Seydi Dogan, Songul Duman, Afsoun Ashkhasi, and
  E.~Fahri Keskenler.
\newblock Structural characterizations and optical properties of inse and
  inse:ag semiconductors grown by bridgman/stockbarger technique.
\newblock {\em Physica E: Low-dimensional Systems and Nanostructures},
  64:106--111, 2014.

\bibitem{Temp-Bandgap-InSe-2020}
M.~Isik and N.M. Gasanly.
\newblock Temperature-tuned band gap characteristics of inse layered
  semiconductor single crystals.
\newblock {\em Materials Science in Semiconductor Processing}, 107:104862,
  2020.

\bibitem{VARSHNI_1967}
Y.P. Varshni.
\newblock Temperature dependence of the energy gap in semiconductors.
\newblock {\em Physica}, 34(1):149--154, 1967.

\bibitem{CanJPhys_1984}
A.~Manoogian and J.~C. Woolley.
\newblock Temperature dependence of the energy gap in semiconductors.
\newblock {\em Canadian Journal of Physics}, 62(3):285--287, 1984.

\bibitem{InSeDebyeTempBook}
O.~Madelung, U.~R{\"o}ssler, and M.~Schulz.
\newblock Indium selenide (inse) debye temperature, heat capacity, density,
  melting point: Datasheet from landolt-b{\"o}rnstein - group iii condensed
  matter {\textperiodcentered} volume 41c: ``non-tetrahedrally bonded elements
  and binary compounds i'' in springermaterials
  (https://doi.org/10.1007/10681727{\_}514).

\bibitem{InSeDebye-Tyurin2007}
A.~V. Tyurin, K.~S. Gavrichev, and V.~P. Zlomanov.
\newblock Low-temperature heat capacity and thermodynamic properties of inse.
\newblock {\em Inorganic Materials}, 43(9):921--925, Sep 2007.

\bibitem{UllrichAIP2017}
Bruno Ullrich, Mithun Bhowmick, and Haowen Xi.
\newblock Relation between debye temperature and energy band gap of
  semiconductors.
\newblock {\em AIP Advances}, 7(4):045109, 2017.

\end{thebibliography}

\end{document}